\documentclass[aps,prd,twocolumn,showpacs,amsmath,amssymb,nofootinbib,eqsecnum, preprintnumbers]{revtex4-1}
\usepackage{graphicx}
\usepackage{epstopdf}
\usepackage{epsfig}
\usepackage{xcolor}
\usepackage{setspace}
\usepackage{caption}
\def\appendix{{\newpage\section*{Appendix}}\let\appendix\section%
        {\setcounter{section}{0}
        \gdef\thesection{\Alph{section}}}\section}
\newcommand{\be}{\begin{equation}}
\newcommand{\ee}{\end{equation}}
\newcommand{\bear}{\begin{eqnarray}}
\newcommand{\eear}{\end{eqnarray}}
\newcommand{\ba}{\begin{array}}
\newcommand{\ea}{\end{array}}

\begin{document}
\title{Configuration entropy and confinement-deconfinement transition\\ in higher-dimensional hard wall model}
\author{Chong Oh Lee}
\email{cohlee@gmail.com}
\affiliation{College of Liberal Arts, Wonkwang University, Iksan, Jeonbuk 54538, Republic of Korea}

\begin{abstract}
We consider a higher-dimensional hard wall model with an infrared (IR) cut-off in asymptotically AdS space and investigate its thermodynamics via the holographic renormalization method.
We find a relation between the confinement temperature and the IR cut-off for any dimension.
It is also shown that the entropy of $p$-branes with the number of coincident branes (the number of the gauge group) $N$ jumps from leading order in $\cal O$($N^0$) at the confining low temperature phase
to $\cal O$($N^{\frac{p+1}{2}}$) at the deconfining high temperature phase like $D3$-branes ($p=3$) case.
On the other hand, we calculate the configuration entropy (CE) of various magnitudes of an inverse temperature at an given IR cut-off scale.
It is shown that as the inverse temperature grows up, the CE above the critical temperature decreases and AdS black hole (BH) is stable  while it below the critical temperature is constant and thermal AdS (ThAdS) is stable.
In particular, we also find that the CE below the critical temperature becomes constant and its magnitude increases as a dimension of AdS space increases.
\end{abstract}

\maketitle
\newpage

\section{Introduction}

The AdS/CFT duality has been originally based on the holographic principle which states that the description of
a volume of space is able to be thought of as encoded on a boundary of such a region and conjectured to be a relationship between a gravitational theory in the bulk and a conformal field theory in the boundary~\cite{Maldacena:1997re,Gubser:1998bc,Witten:1998qj}.
In particular, it has been suggested that BH thermodynamics in AdS is the Hawking-Page transition between large BH and ThAdS,
which in the CFT is dual to the confinement-deconfinement transition on a sphere~\cite{Witten:1998zw}.

The Dirichlet branes (D-branes) have been defined by Dirichlet  boundary conditions in string theory, which has opened up a new window to explore the black hole entropy~\cite{Polchinski:1995mt}. In extremal five-dimensional black hole, they have shown that its horizon area is non-vanishing and the Bekenstein-Hawking entropy is derived by counting the degeneracy of Bogomolny-Prasad-Sommerfield soliton bound states~\cite{Strominger:1996sh}. It has been extensively studied for extremal five-dimensional rotating charged BH~\cite{Breckenridge:1996is}, non-extremal five-dimensional BH~\cite{Callan:1996dv} and non-extremal six-dimensional black string~\cite{Horowitz:1996fn}. The entropies of $Dp$-branes have been investigated~\cite{Gubser:1996de}.

The AdS/QCD hard wall model has been suggested in search in order to obtain the holographic dual theory from QCD~\cite{Erlich:2005qh, Karch:2006pv} via the holographic renormalization method~\cite{Bianchi:2001kw, Skenderis:2002wp}. They have shown that it is able to describe the conﬁnement-deconﬁnement transition for the gauge theory from BH thermodynamics in AdS~\cite{Herzog:2006ra, BallonBayona:2007vp}.
Recently, it has been found that the above result is consistent with that of the analysis of stability through employing the CE
in hard wall model~\cite{Braga:2020opg} and in soft wall model~\cite{Braga:2021zyi}.
Furthermore the AdS/CFT correspondence holds in different dimensions~\cite{Witten:1998zw} but their works~\cite{Braga:2020opg, Braga:2021zyi} focus on the case of the five-dimensional AdS.
Thus, it is intriguing that the issue is generalized to a variety of AdS space.

The paper is organized as follows: In the next section we briefly review the holographic renormalization method for the $d$-dimensional gravity action with the negative cosmological constant
and investigate the relationship between the conﬁnement temperature and the IR cut-oﬀ for the hard wall model at finite temperature. In particular we calculate the entropy of $p$-branes at  the conﬁning temperature phase and the deconﬁning temperature phase. In the next section, we also introduce the CE and explore thermodynamic instability for the hard wall model. In the last section we give our discussion.

\section{Thermodynamics for the Hard wall model}
The $d$-dimensional gravity action $I$ with the negative cosmological constant~\cite{Balasubramanian:1999re, Kraus:1999di} is written as
\bear\label{I}
I=I_{\rm B}+I_{\rm \partial B}+I_{\rm ct}
\eear
where the bulk action $I_{\rm B}$ is given as
\bear\label{IB1}
I_{\rm B}=-\frac{1}{16\pi G_d}\int_{\cal M} d^dx\sqrt{-g}({\cal R}-2\Lambda).
\eear
Here $G_d$, $\cal R$, and $\Lambda$ denote the $d$-dimensional Newton constant, Ricci scalar, and negative cosmological constant $\Lambda=-\frac{(d-1)(d-2)}{2L^2}$.
The boundary action $I_{\rm \partial B}$ is
\bear\label{Ib}
I_{\rm \partial B}=-\frac{1}{8\pi G_d}\int_{\partial \cal M} d^{d-1}x\sqrt{-\gamma}\,\Theta,
\eear
which is introduced to obtain equations of motion well behaved at the boundary of  the action
where $\gamma$ is the determinant of the metric of the boundary $\gamma_{ab}$ and
$\Theta$ is the trace of extrinsic curvature. Then the boundary energy-momentum tensor becomes
\bear
\frac{2}{\sqrt{-\gamma}}\frac{\delta I_{\rm \partial B}}{\delta\gamma^{ab}}=\Theta_{ab}-\gamma_{ab}\Theta.
\eear
The counterterm action $I_{\rm ct}$ is added to the action to remove the divergence appearing as the boundary goes
to infinity
\bear\label{CII}
I_{\rm ct}&=&\frac{1}{8\pi G_d}\int_{\partial \cal M} d^{d-1}x\sqrt{-\gamma}\left\{\frac{d-2}{L}\right.\nonumber\\
&&+\frac{LR}{2(d-3)}{\cal F}(d-4)+\frac{L^3}{2(d-3)^2(d-5)}\bigg(R_{ab}R^{ab}\nonumber\\
&&\left.\left.-\frac{d-1}{4(d-2)}R^2\right){\cal F}(d-6)+\cdots\right\},
\eear
where $R$ is the boundary Ricci scalar which only depends on the induced
metric $\gamma_{ab}$ and ${\cal F}(d)$ is step function, 1 when $d\geq0$, 0 otherwise.

ThAdS with the line element becomes
\bear
ds^2=\frac{r^2}{L^2}(-dt^2+dx_i^2)+\frac{L^2}{r^2}dr^2,
\eear
and substituting with $r=L^2/z$
\bear\label{AdS}
ds^2=\frac{L^2}{z^2}(-dt^2+dx_i^2+dz^2).
\eear
 The Schwarzschild BH in AdS space is given as
\bear
ds^2&=&\frac{r^2}{L^2}\left[-\left(1-\frac{r_h^{d-1}}{r^{d-1}}\right)dt^2+dx_i^2\right]\nonumber\\
&&+\frac{L^2}{r^2}\left(1-\frac{r_h^{d-1}}{r^{d-1}}\right)^{-1}dr^2,
\eear
and substituting with $r=L^2/z$
\bear\label{BH}
ds^2&=&\frac{L^2}{z^2}\left[-\left(1-\frac{z^{d-1}}{z_h^{d-1}}\right)dt^2+dx_i^2\right.\nonumber\\
&&\left.+\left(1-\frac{z^{d-1}}{z_h^{d-1}}\right)^{-1}dz^2\right],
\eear
which leads to the Hawking temperature of AdS BH $T_{H}=(d-1)/(4\pi z_h)$ $(T_{H}\equiv 1/\beta)$
where $z_h$ is AdS BH horizon radius.

The bulk action $I_{\rm B}$ (\ref{IB1}) for both geometries (\ref{AdS}) and (\ref{BH}) becomes
\bear\label{IB2}
I_{\rm B}=\frac{d-1}{8\pi G_d L^2}\int d^dx\sqrt{-g},
\eear
which  has the interval $0< z \leq z_0$ via  the IR cut-off in the hard wall model~\cite{Polchinski:2001tt, Boschi-Filho:2002wdj, Boschi-Filho:2002xih}.
Here the inverse of $z_0$ may be interpreted as  a IR energy cut-oﬀ in the dual gauge theory side.
Furthermore the bulk action (\ref{IB2}) becomes singular as the coordinate $z$ goes to zero.
We may restrict ourselves to the regularized action densities, which for AdS and AdS BH are written as
\bear
{\cal I}_{\rm B}^{\rm AdS}&=&\frac{(d-1)L^{d-2}}{k_d^2}\int_{0}^{\beta'}d\tau\int_{\epsilon}^{z_0}\frac{dz}{z^d}\nonumber\\
&=&\frac{L^{d-2}\beta'}{k_d^2}\left(\frac{1}{\epsilon^{d-1}}-\frac{1}{z_0^{d-1}}\right),\\
{\cal I}_{\rm B}^{\rm BH}&=&\frac{(d-1)L^{d-2}}{k_d^2}\int_{0}^{\beta}d\tau\int_{\epsilon}^{z_{\rm m}}\frac{dz}{z^d}\nonumber\\
&=&\frac{L^{d-2}\beta}{k_d^2}\left(\frac{1}{\epsilon^{d-1}}-\frac{1}{z_{\rm m}^{d-1}}\right),
\eear
with $k_d^2=8\pi G_d$ and $\beta'=\pi z_h\sqrt{1-(\epsilon/z_h)^{d-1}}$~\cite{Herzog:2006ra}.
Here $\epsilon$ is the ultraviolet regulator and $z_{\rm m}\equiv {\rm min}(z_0,z_h)$ denotes the minimum value of $z_0$ and $z_h$.

From Eqs. (\ref{Ib}), (\ref{AdS}) and (\ref{BH}), the boundary action densities are obtained as
\bear
{\cal I}_{\rm \partial B}^{\rm AdS}&=&-\frac{(d-1)L^{d-2}\beta'}{k_d^2}\frac{1}{\epsilon^{d-1}},\\
{\cal I}_{\rm \partial B}^{\rm BH}&=&-\frac{(d-1)L^{d-2}\beta}{k_d^2}\left(\frac{1}{\epsilon^{d-1}}-\frac{1}{2z_h^{d-1}}\right).
\eear

The counterterm action $I_{\rm ct}$ (\ref{CII}) for both geometries (\ref{AdS}) and (\ref{BH}) results in a simple form
\bear\label{CI}
I_{\rm ct}&=&\frac{d-2}{k_d^2 L}\int_{\partial \cal M} d^{d-1}x\sqrt{-\gamma},
\eear
which leads to the counterterm action densities
\bear
{\cal I}_{\rm ct}^{\rm AdS}&=&\frac{(d-2)L^{d-2}\beta'}{k_d^2}\frac{1}{\epsilon^{d-1}},\\
{\cal I}_{\rm ct}^{\rm BH}&=&\frac{(d-2)L^{d-2}\beta}{k_d^2}\left(\frac{1}{\epsilon^{d-1}}-\frac{1}{2z_h^{d-1}}\right).
\eear

Thus, from the $d$-dimensional gravity action $I$ (\ref{I}), the action densities become
\bear
{\cal I}_{\rm AdS}&=&-\frac{L^{d-2}\beta'}{k_d^2}\frac{1}{z_0^{d-1}},\\
{\cal I}_{\rm BH}&=&-\frac{L^{d-2}\beta}{k_d^2}\left(\frac{1}{z_{\rm m}^{d-1}}-\frac{1}{2z_h^{d-1}}\right).
\eear

The difference between the action densities is defined as
\bear
\Delta {\cal I}\equiv\lim_{\epsilon\rightarrow 0}[{\cal I}_{\rm BH}-{\cal I}_{\rm AdS}],
\eear
and
\bear
\hspace{-20pt}\Delta {\cal I} &=&\frac{4\pi L^{d-2} z_h}{(d-1)k_d^2}\frac{1}{2z_{h}^{d-1}}~{\rm for}~z_0<z_h,\\
\hspace{-20pt}\Delta {\cal I} &=&\frac{4\pi L^{d-2} z_h}{(d-1)k_d^2}\left(\frac{1}{z_{0}^{d-1}}-\frac{1}{2z_{h}^{d-1}}\right)~{\rm for}~z_0>z_h,
\eear
which leads to the critical temperature
\bear\label{Tc}
T_{c}=\frac{2^{\frac{1}{d-1}}}{z_0 \pi}.
\eear

After employing $\log Z=-I$ as the partition function, one can get the following  thermal relation $F=-T\log Z = T I$.
The free energies for AdS case and AdS BH are given as
\bear
F_{\rm AdS}=-\frac{L^{d-2}}{k_d^2 z_0^{d-1}}=-\frac{L^{d-2}}{8\pi G_d z_0^{d-1}},
\eear
\bear
F_{\rm BH}=-\frac{\pi^{d-1}L^{d-2}}{2k_d^2}T^{d-1}=-\frac{(\pi L)^{d-2}}{16 G_d}T^{d-1},
\eear
which become respectively
\bear
F_{\rm AdS}&=&-\frac{L^{p}}{k_{p+2}^2 z_0^{p+1}}\nonumber\\
&=&-\frac{L^{p}}{8\pi G_{p+1} z_0^{p+1}}~{\rm for}~(T_c >T),
\eear
\bear\label{pFE}
F_{\rm BH}&=&-\frac{\pi^{p+1}L^{p}}{2k_{p+2}^2}T^{p+1}\nonumber\\
&=&-\frac{(\pi L)^{p}}{16 G_{p+2}}T^{p+1}~{\rm for}~(T_c <T),
\eear
where $d=p+2$ and $p$ is dimension of brane.
\begin{center}
\begin{table*}[hbt!]
\begin{center}
\caption{The CE of AdS BH and ThAdS for the various dimensions.}
\end{center}
{\small
\begin{tabular}{ |c|c|c|c|c|c|c|c| }
\hline
$z_h$    &$d=4$      &$d=5$      &$d=6$      &$d=7$      &$d=8$      &$d=9$      &Type\\
\hline
0.01   &17.95581731&17.95595568&17.95597083&17.95597251&17.95597269&17.95597271&AdS BH\\
0.05   &16.34637939&16.34651777&16.34653292&16.34653460&16.34653469&16.34653473&AdS BH\\
0.1    &15.65323222&15.65337059&15.65338574&15.65338741&15.65338760&15.65338762&AdS BH\\
0.5    &14.55461993&14.55475830&14.55477345&14.55477513&14.55477521&14.55477533&AdS BH\\
0.7    &14.04379431&14.04393268&14.04394783&14.04394950&14.04394969&14.04394971&AdS BH\\
$z_c$  &13.70732207&13.70746044&13.70747559&13.70747726&13.70747745&13.70747747&AdS BH\\
\hline
$z_h>z_c$&13.56214063&13.50450336&13.46985271&13.44674434&13.43023742&13.41785715&ThAdS\\
\hline
\end{tabular}
\caption*{Here, $z_c$ denotes the horizon radius at the critical temperature.}
}
\end{table*}
\end{center}

The supergravity solution reduces to $p+2$-dimensional AdS space ($AdS_{p+2}$) times spheres
(i.e. $AdS_5 \times S_5$ for $D3$-branes). The entropy of the non-dilatonic near-extremal $p$-branes
is calculated from the leading order supergravity solution. The free energy of CFT side~\cite{Klebanov:1996un} is expected to have the form
\bear\label{FE}
F=-\alpha_{p} N^{\frac{p+1}{2}}T^{p+1},
\eear
where $\alpha_{p}$ is a constant of order unity.

In the $D3$-branes case (total dimension $D = 10$ and $p = 3$, $AdS_5 \times S^5$), the free energy (\ref{pFE}) is consistent with that in~\cite{Witten:1998zw}
\bear
F_{\rm BH}=-\frac{\pi^2}{8}N^2T^4,
\eear
where the five-dimensional Newton constant $G_5$ and AdS radius $L$ are
\bear\label{G5}
G_5=\frac{8\pi^3 g^2 \alpha'^4}{L^5}~{\rm and}~L^4=4\pi g N \alpha'^2.
\eear
In the $M5$-branes case (total dimension $D = 11$ and $p = 5$, $AdS_7 \times S^4$), the free energy (\ref{pFE}) becomes the free energy of AdS BH~\cite{Klebanov:1996un}
\bear
F_{\rm BH}=-\frac{2^6 \pi^3}{3^7}N^3T^6,
\eear
where the seven-dimensional Newton constant $G_7$ and AdS radius $L$ are
\bear\label{G7}
G_7=\frac{3^7 \pi^2 L^5}{2^{10}N^3}~{\rm and}~L^9=G_7\frac{N^{3}}{2^4 \pi^4}.
\eear
In the $M2$-branes case (total dimension $D = 11$ and $p = 2$, $AdS_4 \times S^7$), the free energy (\ref{pFE}) becomes the free energy of AdS BH~\cite{Klebanov:1996un}
\bear
F_{\rm BH}=-\frac{2^{7/2}\pi^2}{3^4}N^{3/2}T^3,
\eear
where the four-dimensional Newton constant $G_4$ and AdS radius $L$ are
\bear\label{G4}
G_4=\frac{3^4 L^2}{2^{15/2}N^3}~{\rm and}~L^9=G_4\frac{2^{7/2}N^{3/2}}{\pi^4}.
\eear

On the other hand, the expectation value of the energy is given as
\bear\label{E}
<E>=-\partial_{\beta}\log Z=\partial_{\beta} I\sim -F.
\eear
After employing the Gibbs-Duhem relation $S=\beta <E>+\log Z=\beta<E>-I$,
the entropies for AdS case and AdS BH are given as
\bear
S_{\rm AdS}=0\sim N^0~{\rm for}~(T_c >T),
\eear
\bear\label{EN}
S_{\rm BH}&=&\frac{(d-1)(\pi L)^{d-2}}{16 G_d}T^{d-2},\\\label{EN1}
&=&\frac{(p+1)(\pi L)^{p}}{16 G_{p+1}}T^{p}~{\rm for}~(T_c <T),
\eear
where after employing the relation~(\ref{E}) between the expectation value of the energy and the free energy, and substituting
the Gibbs-Duhem relation with the free energy~(\ref{FE}), we can read
\bear\label{EN2}
S_{\rm BH}\sim N^{\frac{p+1}{2}}T^{p}\sim N^{\frac{p+1}{2}}.
\eear
It is shown that the entropy of $p$-branes jumps from $\cal O$($N^0$) at the confining low temperature phase to $\cal O$($N^{\frac{p+1}{2}}$) at the deconfining high temperature phase
like $D3$-branes ($p=3$) case, which is consistent with the jump in the entropy describing
the change of degrees of freedom in the confinement-deconfinement phase transition of QCD~\cite{Herzog:2006ra, BallonBayona:2007vp}.

Now, we will check the above relation through explicitly calculating for specific cases. Adopting the five-dimensional Newton constant $G_7$ (\ref{G5}) and AdS radius $L$ (\ref{G5}),
the entropy (\ref{EN1}) in the $D3$-branes case is given as
\bear
S_{\rm BH}=\frac{1}{2}\pi^2 N^2 T^3\sim N^2T^3\sim N^2~{\rm for}~p=3.
\eear
In the same way, employing the seven-dimensional Newton constant $G_7$ (\ref{G7}) and AdS radius $L$ (\ref{G7}) in the $M5$-branes case,
and the four-dimensional Newton constant $G_4$ (\ref{G4}) and AdS radius $L$ (\ref{G4}) in the $M2$-branes case,
the entropies (\ref{EN1}) are respectively
\bear
S_{\rm BH}=\frac{5}{128}\pi^5 N^3 T^9\sim N^3T^9\sim N^3~{\rm for}~p=5,
\eear
\bear
\hspace{-10pt} S_{\rm BH}=5\sqrt{2}\pi^5 N^{\frac{3}{2}} T^9\sim N^{\frac{3}{2}}T^9\sim N^{\frac{3}{2}}~{\rm for}~p=2,
\eear
which correctly match with the above result $S_{\rm BH}\sim N^{\frac{p+1}{2}}$ (\ref{EN2}).

Finally after employing thermal relation $C=-\beta \partial_\beta S$, the specific heat for  AdS case and AdS BH are given as
\bear
C_{\rm AdS}=0~{\rm for}~(T_c >T),
\eear
\bear
C_{\rm BH}&=&\frac{(d-1)(d-2)(\pi L)^{d-2}}{16 G_d}T^{d-2},\\
&=&\frac{p(p+1)(\pi L)^{p}}{16 G_{p+2}}T^{p}~{\rm for}~(T_c <T).
\eear
In particular, since AdS BH for $T_c <T$ always has positive specific heat it is thermodynamically stable.

\begin{figure*}[hbt!]
\begin{center}
\includegraphics[width=7cm]{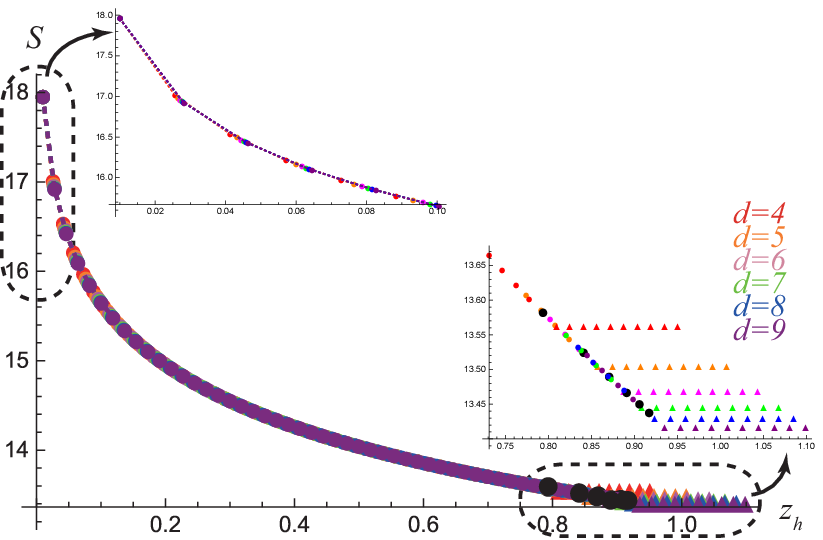}~~~~~
\includegraphics[width=7cm]{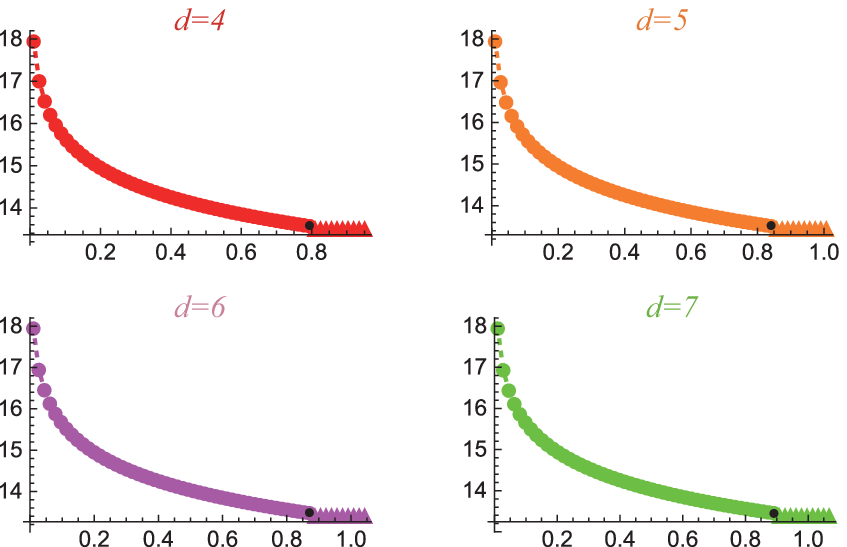}
\caption{Plot of the $d$-dimensional CE $S$ as a function of the horizon radius $z_h$ for $z_0=1$, $k_d=1$ $(8\pi G_d=1)$, and $L=1$.
The black points denote the horizon radius $z_c$ at the critical temperatures for each dimension.}
\end{center}
\end{figure*}

\section{CE of ThAdS and AdS BHs}
After employing thermal relation $\partial_{\beta} I=M$, the mass for ThAdS is obtain as
\bear\label{AdSmass}
M_{\rm AdS}=-\frac{L^{d-2}}{k_d^2z_0^{d-1}},
\eear
and the mass for AdS BH becomes
\bear\label{BHmass}
M_{\rm BH}=\frac{(d-2)L^{d-2}}{2k_d^2z_h^{d-1}}.
\eear
After adopting the relation between the mass and the density $\int_0^{z_{\rm m}} dz \rho(z)=M$,
one can read the regularized density from the ThAdS mass~\cite{Braga:2020opg}
\bear
\rho_{\rm AdS}=\frac{(d-1)L^{d-2}}{k_d^2}\frac{1}{z^d}\cos\left(\frac{2\pi \epsilon^{d-1}}{z^{d-1}}\right),
\eear
which leads to
\bear
\int_{\epsilon}^{z_0}\rho_{\rm AdS}dz=-\frac{L^{d-2}}{2\pi k_d^2\epsilon^{d-1}}\sin\left(\frac{2\pi\epsilon^{d-1}}{z_0^{d-1}}\right),
\eear
which reduces to the mass of ThAdS $M_{\rm AdS}$ (\ref{AdSmass}) in the limit of $\epsilon\rightarrow0$.
The regularized density of the AdS BH mass is written as
\bear
\rho_{\rm BH}=-\frac{(d-1)(d-2)L^{d-2}}{2k_d^2}\frac{1}{z^d}\cos\left(\frac{2\pi \epsilon^{d-1}}{z^{d-1}}\right),
\eear
and
\bear
\int_{\epsilon}^{z_h}\rho_{\rm BH}dz=-\frac{(d-2)L^{d-2}}{4\pi k_d^2\epsilon^{d-1}}\sin\left(\frac{2\pi\epsilon^{d-1}}{z_h^{d-1}}\right),
\eear
which becomes the mass of AdS BH $M_{\rm BH}$~(\ref{BHmass}) in the limit of $\epsilon\rightarrow0$.

When we consider the energy density $\rho(z)$ as the function of the position $z$ in $d$-dimensional space and its Fourier transforms $\rho(k)$
is written as
\bear
\rho(k)=\left(\frac{1}{\sqrt{2\pi}}\right)^d\int \rho(z) e^{-ik\cdot z}d^dx,
\eear
and the modal fraction is defined as
\bear
{\cal F}(k)=\frac{|\rho(k)|^2}{\int|\rho(k)|^2d^dk }.
\eear
One may define the CE~\cite{Gleiser:2012tu} as
\bear
S[{\cal F}]=-\int_{-\infty}^{\infty}{\cal F}(k)\log[{\cal F}(k)]d^dk.
\eear

As shown in Table I, we calculate the entropies for the variety of the dimension of AdS space from $d=4$ to $d=9$ and plot them as you see in Fig.1.
Since the temperature $T$ is given as $T=(d-1)/(4 \pi z_h)$, $z_h$ is inversely proportional to $T$. It is shown that  as the inverse temperature increases, the CE above the critical temperature~(\ref{Tc}) decreases and AdS BH becomes stable while it below the critical temperature~(\ref{Tc}) is constant and ThAdS is stable.
Furthermore, when there is AdS BH above the critical temperature in the stable phase, the CE increases as the temperature increases, which correctly coincides with the expected result that the evaporation with arising from emitting AdS BH radiation increases faster at higher temperatures. These results are similar to that in~\cite{Braga:2020opg}.

\section*{Discussion}
For any dimension we considered the hard wall model in the context of the AdS/CFT duality and explicitly presented the critical temperature $T_c=2^{\frac{1}{d-1}}/(z_0\pi)$~(\ref{Tc}) via the holographic renormalization method.
Then the entropy of $p$-branes jumps from leading order in $\cal O$($N^0$) at the confining low temperature phase below the critical temperature to $\cal O$($N^{\frac{p+1}{2}}$) at the deconfining high temperature phase above the critical temperature like $D3$-branes ($p=3$) case~\cite{BallonBayona:2007vp}. It was found for the hard wall model that AdS BH above the critical temperature~(\ref{Tc}) becomes stable while ThAdS below the critical temperature~(\ref{Tc}) is stable.
We also obtained the same results through thermodynamic analysis of calculating the CE.
It was shown that it is well held to describe the conﬁnement-deconﬁnement transition for the gauge theory from BH thermodynamics in AdS for the higher-dimensional hard wall model beyond the five-dimensional case~\cite{BallonBayona:2007vp}.
In particular, we found that as the temperature grows up, the CE has constant until the critical temperature while it decreases beyond the critical temperature. This result is consistent with that in~\cite{Braga:2020opg}.

Our research here focused on the higher-dimensional hard wall model. Thus, it will be intriguing that the issue is generalized to the higher-dimensional soft wall model.  Adopting the CE, one can explore not only its instability but also conventional hadrons and exotica ones~\cite{Colangelo:2018mrt}.

On the other hand, it is possible to investigate the stability of the AdS black string solution via the CE.
It provides a new window in order to check the Gubser-Mitra conjecture~\cite{Gubser:2000mm,Gubser:2000ec}. We hope that such kind issues will be carried out in near the future.

\section*{Declaration of Competing Interest}
The authors declare the following financial interests/personal relationships which may be considered as potential competing interests: Chong Oh Lee reports financial support was provided by Ministry of Education, Science and Technology (NRF-2018R1D1A1B07049451). Chong Oh Lee reports a relationship with Wonkwang University that includes: employment and funding grants.

\section*{Acknowledgements}
{This paper was supported by Wonkwang University in 2021.}

\end{document}